\documentclass[reprint,
superscriptaddress,
groupedaddress,
showpacs,preprintnumbers,
nobibnotes,
amsmath,amssymb,
aps,
pra,
floatfix,
]{revtex4-1}

\usepackage{graphicx}% Include figure files
\usepackage{dcolumn}% Align table columns on decimal point
\usepackage{bm}% bold math
\usepackage{epstopdf}
\usepackage{float}
\usepackage[caption = false]{subfig}
\usepackage{array}
\usepackage{braket}
\usepackage{physics}
\usepackage{graphicx}
\usepackage{cancel}
\usepackage{ulem}
\begin{document}

%\preprint{APS}

\title{Direct Dissociative Recombination and Ion-pair Formation of $\mathrm{HeH}^+$ Isotopologues}

\author{Sifiso M. Nkambule}
\email{snkambule@uniswa.sz}
 %\affiliation{Department of Physics, Stockholm University, Albanova University Center, SE-10691 Stockholm, Sweden.}
\affiliation{Department of Physics, University of Eswatini, Kwaluseni, M201, Eswatini.} 
\author{Malibongwe Tsabedze}
%\email{snkambule@uniswa.sz}
 %\affiliation{Department of Physics, Stockholm University, Albanova University Center, SE-10691 Stockholm, Sweden.}
\affiliation{Department of Physics, University of Eswatini, Kwaluseni, M201, Eswatini.} 
\author{Oscar N. Mabuza}%
% \email{zwanesibusiso@live.com}
\affiliation{Department of Physics, William Pitcher College , Manzini, M200, Eswatini.}%
\affiliation{Department of Physics, University of Eswatini, Kwaluseni, M201, Eswatini.} 
\author{Mbuso K. Matfunjwa}%
% \email{zwanesibusiso@live.com}
% 
\affiliation{Department of Physics and Astronomy, University of Nebraska, Lincoln, Nebraska 68588-0299, USA}


\affiliation{Department of Physics, University of Eswatini, Kwaluseni, M201, Eswatini.}

\date{\today}% It is always \today, today,

\begin{abstract}
Direct dissociative recombination and resonant ion-pair formation reactions of $\mathrm{HeH^+}$ are theoretically studied using time-dependant wave-packets methods. The wave packets are propagated on potential energy curves that are in either the adiabatic representation or the diabatic representation. The reaction cross sections  are computed for collisions of different isotopes of $\mathrm{He}$ and $\mathrm{H}$. The reactions are modeled in the collision energy range 0 eV to 50 eV. Final states distributions are also investigated for the $\mathrm{^4HeH^+}$ dissociative recombination reaction, showing dominance of contribution  from states of $^2\Sigma$ symmetry in the diabatic representation. In the adiabatic representation, the $^2\Pi$ and $^2\Delta$ states dominate at lower collision energies. The resonant ion-pair formation reaction is investigated using two sets of representation for the potential energy curve of the ion-pair state. The results are compared with available experimental and other theoretical results. The present model yields a reaction cross section that is larger than previous results.

\end{abstract}

%\pacs{34.70.+e, 31.50.Df, 31.50.Gh, 82.20.Bc, 82.20.Ej}% PACS, the Physics and Astronomy
%Use showkeys class option if keyword
%\footnote{* }                              %display desired
\maketitle

%\tableofcontents

\section{Introduction}
\label{intro}


Dissociative recombination (DR) and resonant ion-pair (RIP) formation  processes are amongst some of the most important fundamental chemical processes that modify the charge and energy balance in low-temperature plasmas. DR is a process whereby a molecular cation collides with and captures an electron. This results in the formation of a resonant neutral molecule, in an excited repulsive state which then dissociates into neutral fragments. In the RIP formation process, on the other hand, the resulting fragments consist of an ion-pair. In order to model the plasma environments correctly, a more detailed understanding of the driving mechanisms and reaction cross sections of these processes is needed.

There are two mechanisms in which the DR process proceeds. Firstly, the DR process can take place when an incident electron is captured into a repulsive electronic state of the neutral molecule and the nuclei breaks down in this potential. Such a molecule is stabilized against the competing autoionization process when the internuclear separation is such that the potential energy curve of the neutral lies below the ionic ground-state potential energy curve. This is known as the direct mechanism of the DR reaction. On the other hand, the indirect mechanism for DR occurs when the electron is captured into a vibrationally excited molecular Rydberg state, of the neutral molecule. It then predissociates with the repulsive state. It has been proposed that of the two mechanisms, the direct DR mechanism is normally the most dominant process~\cite{Bardsley2}. However, Sarpal \textit{et al}~\cite{Sarpal94} pointed out that interference between the two processes may generally result in complicated resonance structure in the cross section. For the $\mathrm{HeH}$ system, the covalent resonat states potential energy curves do not cross the potential energy curve of $\mathrm{HeH^+}$~\cite{Larson2014}. However, it has been previously discussed~\cite{Orel95} that direct DR is still possible, while autoinization at short internuclear distances, is a possibility as well.

In early universe gas evolution, the $\mathrm{HeH^+}$ ion is believed to have played a very important role in gas cooling~\cite{Lepp02}. It is also believed that this ion is one of the key ingrediants neccessary  in studies of the understanding the chemistry of interstellar plasmas~\cite{black78,stromholm}. There were previously some difficulty in detecting the ion in the interstellar medium~\cite{moorhead} and this seemed to contradict the predictions that $\mathrm{HeH^+}$ could be abundant~\cite{roberge}. However, it was later found that these predictions were neglecting the destruction of $\mathrm{HeH^+}$ by the DR process~\cite{guberman}. This process can be represented as
\begin{equation}
\mathrm{HeH^+}+ \mathrm{e^-}\rightarrow \mathrm{HeH^*}\rightarrow \mathrm{He}+\mathrm{H},
\label{DR1}
\end{equation}
where $\mathrm{HeH^*}$ denotes a neutral, repulsive and unstable resonant state which then breaks down to the fragments, $\mathrm{He}$ and $\mathrm{H}$. There is a possibility of these fragments to be electronically excited. The RIP formation reaction, on the other hand, can be represented as
\begin{equation}
\mathrm{HeH^+}+ \mathrm{e^-}\rightarrow \mathrm{HeH^*}\rightarrow \mathrm{He^+}+\mathrm{H^-}.
\label{RIP1}
\end{equation}
It can be seen that processes~(\ref{DR1}) and~(\ref{RIP1}) only differ in the final fragments formed, on whether they are charged on neutral. They are therefore competing procceses amongst many other processes~\cite{Larson2014}.

In the interstellar environments, $\mathrm{HeH^+}$ is expected to participate in a large variety of other processes, such as the spontaneous emission of low energy photons from the ion. It has been suggested that such a process could be an important factor in primordial star formation~\cite{lepp84}. The collision of an electron with the molecular ion could result in a number of other processes, ranging from elastic scattering to inelastic processes. These processes are vibrational excitation, rotational excitation and dissociative excitation, to name just a few. All these processes are expected to compete with processes ~(\ref{DR1}) and~(\ref{RIP1}). It is speculated that $\mathrm{HeH^+}$ is probably one of the oldest molecular ions in the universe and it had eluded astrophysical observation for decades. It was until recently that Guesten \textit{et al}~\cite{Guesten19} finally reported a positive detection of the ions in the nebula NGC 7027.

When modeling the chemistry of the interstellar medium, an observable abundance of $\mathrm{HeH^+}$ was predicted~\cite{roberge}. Further, it was proposed that the destruction rate of $\mathrm{HeH^+}$ is increased by the DR process with low energy electrons. Numerous theoretical calculations of potential energy curves for $\mathrm{HeH^+}$ and $\mathrm{HeH}$ have failed to show the desired crossing of the ionic ground state by a neutral repulsive state. This for a long time,  led to a belief that the DR reaction rate of $\mathrm{HeH^+}$ is very low. On the contrary, single-pass and storage ring experiments all found and confirmed a large recombination rate~\cite{Stromholm94,Mowat95,Sundstrom94,Yousif89}. Further experimental work was carried out by Seminiak \textit{et al}~\cite{Seminiak96} to measure the final products states distribution formed in DR reaction of $\mathrm{^3HeD^+}$ and $\mathrm{^4HeH^+}$. This study was carried out by using the CRYRING~\cite{Stromholm94} and TSR~\cite{Habs89} heavy ion storage rings. The results obtained were in agreement with earlier theoretical results based in MQDT theory~\cite{guberman}. However, it was found that there was a discrepancy with the theoretical results of Sarpal \textit{et al}~\cite{Sarpal94} which were based on the R-matrix theory.

The potential energy curves used in the current study are those reported by Larson \textit{et al}~\cite{Larson15,Larson16}. Eleven potential energy curves for resonant states of $\mathrm{HeH}$ in the $^2\Sigma^+$ symmetry are computed together with six in the $^2\Pi$ symmetry and six in the $^2\Delta$ symmetry. Previously a theoretical study of the DR reaction of $\mathrm{HeH^+}$ was carried out by Larson \textit{et al}~\cite{Larson98}. In the study only six (two of $^2\Sigma$ symmetry and two of $^2\Pi$ symmetry) states were included. A two-by-two adiabatic to diabatic state transformation~\cite{Mead82} was carried out to obtain the potential energy of the ion-pair state and diabatic electronic energy curves. No rotational couplings between states were considered. In the current study, 23 resonant states are included in both the diabatic and adiabatic represenation. These are resonant states that are above the ground state of $\mathrm{HeH^+}$, but below the excited state. They consist of eleven states of $^2\Sigma$, six states of $^2\Pi$ and six states of $^2\Delta$ symmetries. Rotational couplings between states of different symmetries are also included. A strict diabatization of the 23 coupled states is also carried out in order to study the dissociation dynamics in the diabatic representation.

This paper is arranged as follows; Section~\ref{pec} discusses the potential energy curves and  how the rotational couplings are computed. The nuclear dynamics relevant for the direct capture and dissociation amongst the resonant states of $\mathrm{HeH}$ and the ion-pair, $\mathrm{He^+ + H^-}$ are outlined in section~\ref{nd}. Secton~\ref{dresults} contains the results and discussion of the results. Lastly, a conclusion is given in section~\ref{concl}.

\section{\label{pec}Potential energy curves}

The potential curves for the system that are used in the current study are those that are first reported by some of us in ref~\cite{Larson15}, where the full configuration interaction level of theory is used for the electronic structure calculation. This is then combined with electron scattering calculations that are based in the complex-Kohn variational method~\cite{Kohn1}. The electron scattering calculations give the resonant states energy positions above the potential energy curves of the ground state of the ion. The autoinization widths, $\Gamma(R)$, are also obtained and they are non-zero at short internuclear distances. At large internuclear distances, $\Gamma(R) \rightarrow 0$. The adiabatic potential energy curves and autoinization widths are in figures 1 and 4(a) of ref.~\cite{Larson15} , respectively.

In the adiabatic representation, the resonant states of $^2\Sigma$ symmetry do not preserve their character. The state lowest in energy has an ion-pair character at short internuclear distances, while this character is exhibited by the state highest in energy at large internuclear distances. Avoided crossings are observed whenever there is a switch from ion-pair state to a covalent state~\cite{Larson16}. However, in the diabatic representation the lowest resonant state, at short internucler distances, crosses the other resonant states and dissociates to the ion-pair $\mathrm{He^+ + H^-}$, while the other covalent states dissociates to neutral fragments.

\subsection{Rotational Couplings}

The $\bm{L}$-uncoupling term of the rotational Hamiltonian can be given as~\cite{Lefebre},
\begin{eqnarray}
&-\dfrac{1}{2\mu R^2}\bra{J,S,\Omega\pm 1,\Lambda,\Sigma\pm 1}\bm{J_{\pm}L_{\pm}}\ket{J,S,\Omega,\Lambda,\Sigma}&= \nonumber \\
&-\dfrac{1}{2\mu}\left[J\left(J+1\right)-\Omega\left(\Omega\pm 1\right) \right]^{\frac{1}{2}} \times & \nonumber \\
&\bra{J,S,\Omega\pm 1,\Lambda,\Sigma\pm 1}\bm{L_{\pm}}\ket{J,S,\Omega,\Lambda,\Sigma}&,
\label{lun}
\end{eqnarray}
where $J$, $L$, and $S$ are well-known quantum numbers and $\Omega$, $\Lambda$ and $\Sigma$ are their respective projections onto the molecular axis. $\bm{J}$ and $\bm{L}$ are the total angular momentum and total orbital angular momentum operators, respectively. Here we shall denote the molecular electronic wavefunctions for the $^2\Sigma^+$, $^2\Pi$ and $^2\Delta$ states as $\Phi_{\Sigma}$, $\Phi_{\Pi}$ and $\Phi_{\Delta}$, respectively. Based on the selection rules, the allowed difference between any two symmetries is such that $\Delta \Lambda =\pm 1$. Thus rotational couplings between $\Phi_{\Sigma}$ and $\Phi_{\Delta}$ is zero. Compared to the term $\ell (\ell +1)$, the term $\Omega \left(\Omega +1 \right)$ in eq.~(\ref{lun}) is very small ($\frac{1}{4}$) and thus if neglected, the rotational couplings between $\Phi_{\Sigma}$ and $\Phi_{\Pi}$ states are
\begin{equation}
\dfrac{-\left[ \ell(\ell +1)\right]^{\frac{1}{2}}}{2\mu R^2}\bra{\Phi_{\Pi}}\bm{L_{\pm}}\ket{\Phi_{\Sigma}}.
\end{equation}
$\ell$ are the well-known rotational quantum numbers. From the electronic structure calculations for the electronic states of the $\mathrm{HeH}$ system~\cite{Larson15}, the dominant configurations for covalent states associated with the same asymptotc limit is of the form $(1\sigma)^1(2\sigma)^1(n\lambda)^1$. Thus the electronic states of different symmetries in the system are related by
\begin{eqnarray}
\bra{\Phi_{\Pi}}\bm{L_{\pm}}\ket{\Phi_{\Sigma}}=\bra{(np\pi)^1}\bm{\ell_+}\ket{(np\sigma)^1},
\label{np1}\\
\bra{\Phi_{\Delta}}\bm{L_{\pm}}\ket{\Phi_{\Pi}}=\bra{(np\delta)^1}\bm{\ell_+}\ket{(np\pi)^1},
\label{np2}
\end{eqnarray}
where $\bm{L_+}$ has been decomposed into one-electron operators, $\bm{\ell_+}$. Applying the pure precision approximation~\cite{Hemert91}, eq.~(\ref{np1}) and eq.~(\ref{np2})  reduces to
\begin{equation}
\bra{\Phi_{\Pi}}\bm{L_{\pm}}\ket{\Phi_{\Sigma}}=\sqrt{2}
\end{equation}
 and 
 \begin{equation}
\bra{\Phi_{\Delta}}\bm{L_{\pm}}\ket{\Phi_{\Pi}}=\sqrt{2},
\end{equation}
respectively. For covalent states associated with different asymptotic limits,
\begin{eqnarray}
\bra{\Phi_{\Pi}}\bm{L_{\pm}}\ket{\Phi_{\Sigma}}=0\\
\bra{\Phi_{\Delta}}\bm{L_{\pm}}\ket{\Phi_{\Pi}}=0.
\end{eqnarray}

Based on the above discussion for the rotational couplings, the adiabatic potential energy curves for the system is represented with the electronic potential energies obtained from the electronic structure calculations~\cite{Larson15}. In addition, there are now rotational quantum number dependant off-diagonal elements of the form
\begin{equation}
V_{\Pi,\Sigma}^{a,\ell}(R)=-\dfrac{\sqrt{\ell(\ell +1)}}{2\mu R^2}.
\label{vl}
\end{equation} 
$R$ is the internuclear separation distance. 

The strict adiabatic to diabatic transformation is carried out, with the inclusion of rotational couplings, to obtain diabatic potential energy curves and electronic couplings as previously discusssed by Larson \textit{et al}~\cite{Larson16}.

\section{\label{nd}Nuclear Dyamics}

The study of dynamics for the dissociation are carried out by solving the time-dependant Schr\"odinger equation,
\begin{equation}
i\dfrac{\partial }{\partial t}\Psi(R,t) =\bm{H_T}\Psi(R,t).
\label{tdse}
\end{equation}
By direct integration of eq.~(\ref{tdse}), a wave packet is propagated with the local complex potential. For states $i$ and $j$, the Hamiltonian is of the form
\begin{eqnarray}
H_{Tij} &=&\left(-\dfrac{1}{2\mu}\dfrac{\partial^2}{\partial R^2} + V_i(R)-i\dfrac{1}{2}\Gamma_i(R)-\dfrac{\sqrt{\ell(\ell +1)}}{2\mu R^2} \right)\delta_{ij}\nonumber \\
&+& H_{ij}(R).
\label{hamilt2}
\end{eqnarray}
$V_i(R)$ is the potential energy while $\Gamma_i(R)$ is the resonant state autoinization width, obtained from the electron scattering calculations described in ref~\cite{Larson15}. $V_i(R)$ can either be in an adiabatic or diabatic representation and $\mu$ is the reduced mass. $H_{ij}$ describes the couplings between the states. In the adiabatic reprsentation, this term is known as the second order derivative non-adiabatic coupling,
\begin{equation}
H_{ij}(R)=-\dfrac{1}{2\mu} \bra{\Phi_{i}^{a}(r,R)}\dfrac{\partial^2}{\partial R^2}\ket{\Phi_{j}^{a}(r,R)}.
\label{nonadc}
\end{equation}
$\Phi_{i}^{a}(r,R)$ is the electronic wave-function for state $i$. The subscript `$a$' denoted the adibatic representation, and $r$ denotes electronic coordinates. Non-adiabatic coupling elements for electronic states of the $\mathrm{HeH}$ system are shown as figure 3 in ref.~\cite{Larson15} and figures 4 and 5 in ref.~\cite{Larson16}. 

The diabatic representation is by definition~\cite{Mead1982} a basis where
\begin{equation}
-\dfrac{1}{2\mu} \bra{\Phi_{i}^{d}(r,R)}\dfrac{\partial^2}{\partial R^2}\ket{\Phi_{j}^{d}(r,R)}=0.
\end{equation} In this representation, the term $H_{ij}$ in eq.~\ref{hamilt2} is the electronic couplings between states of identical symmetry  and is given by
\begin{equation}
H_{ij}(R)=\bra{\Phi_{i}^{d}(r,R)}\bm{H}_{el}\ket{\Phi_{j}^{d}(r,R)}.
\label{elecc}
\end{equation}
The subscript `$d$' denotes the diabatic representation and $\bm{H}_{el}$ is the electronic Hamiltonian. Although  $ \Phi_{i}^{a}(r,R)$ are eigen-functions of $\bm{H}_{el}$, $\Phi_{i}^{d}(r,R)$ are not and this result in off-diagonal elements of eq~(\ref{elecc}). It has been discussed previoulsy that the electronic couplings of eq.~(\ref{elecc}) varies smoothly with $R$ compared with the nonadiabatic couplings of eq.~(\ref{nonadc})~\cite{Hedberg14,Nkambule15,Nkambule22}.

The initial condition of the wave packet is given by~\cite{Orel95}
\begin{equation}
\Psi(R,t=0)=\sqrt{\dfrac{\Gamma_i(R)}{2\pi}}\chi_{v=0}(R),
\label{init}
\end{equation}
where $\chi_{v=0}(R)$ is the vibrational wavefunction for the $v=0$ vibrational level of the ground state of $\mathrm{HeH^+}$. This is numerically evaluated using a finite difference method~\cite{Truhlar}, employed to solve the time-independant Schr\"odinger equation for the potential energy curve of the ground state of the ion. The wave packets are then propagated on the potential energy curves, in each representation (adiabatic and  diabatic), using a numerical algorithm based on the Cranck-Nicholson propagation method~\cite{Godberg}, while autoionization is included, as a local complex potential, within the ``boomerang" model~\cite{Herzenberg68,Herzenberg79}.

The contributions to the DR reaction and RIP formation reaction cross section from resonant state $i$ is computed with
\begin{equation}
\sigma_i (E) =\dfrac{2\pi^3}{E}\beta_i|Tr_i(E)|^2,
\end{equation}
where $E$ is the electron scattering energy and $\beta_i$ is the multiplicity ratio for the final and initial state. $Tr_i(E)$ is the transition amplitude~\cite{Gertitshke93} and it is obtained by performing a half Fourier transform of the wave packet at the asymptotic region, $R_{asy}$,
\begin{equation}
Tr_i(E) =\sqrt{\dfrac{K}{2\pi \mu}}\int_{0}^{\infty}\Psi_i(R_{asy},t)e^{iEt}dt.
\end{equation}
$\mu$ is the reduced mass of the system, while $K$  is the wave number associated with the dissociating fragments.

\section{\label{dresults} Results}
\subsection{DR Reaction Cross Section for $\mathrm{^4HeH^+}$}

\begin{figure}[!h]
\includegraphics[scale=0.35,angle=-90]{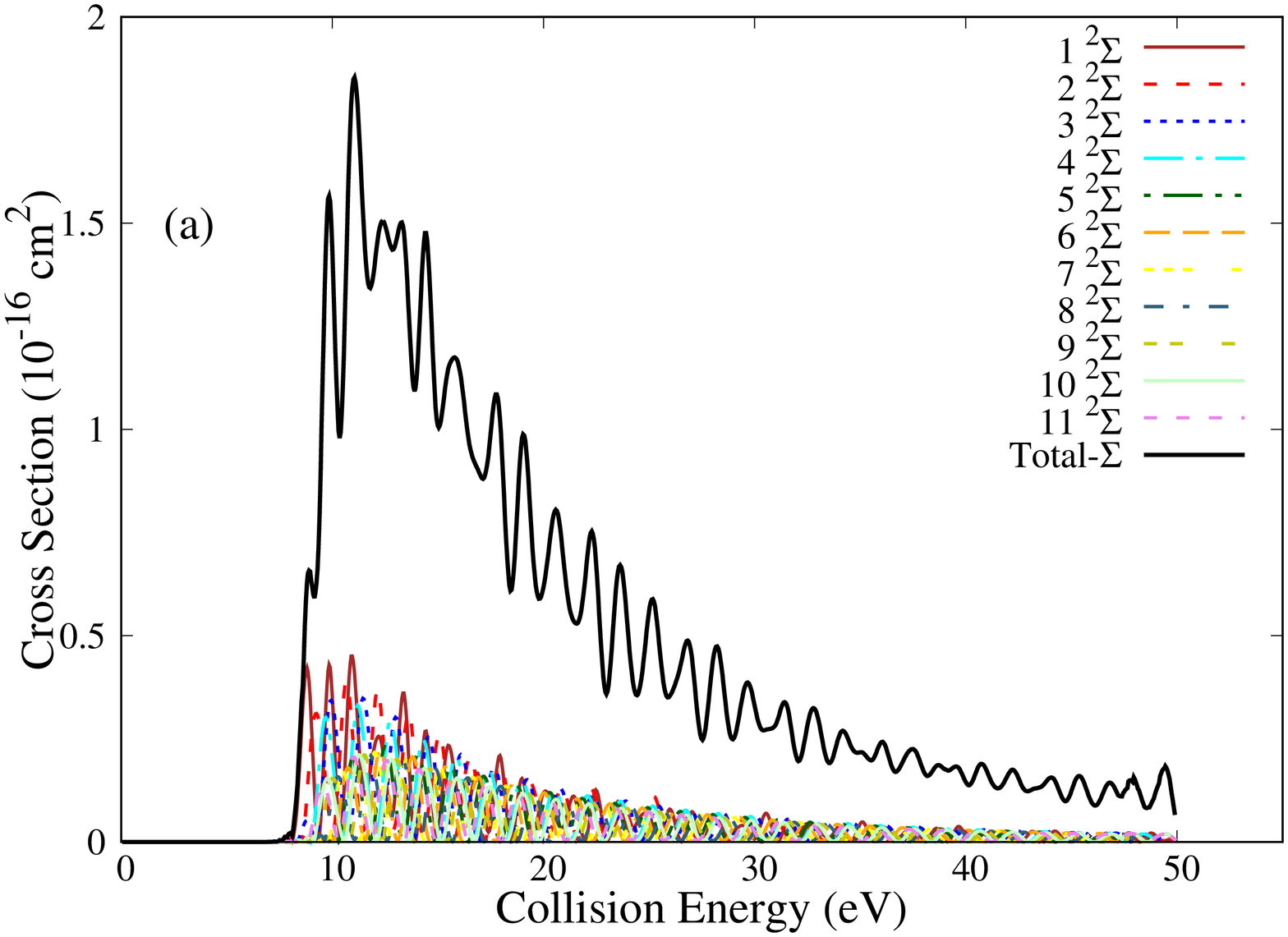}
\includegraphics[scale=0.35,angle=-90]{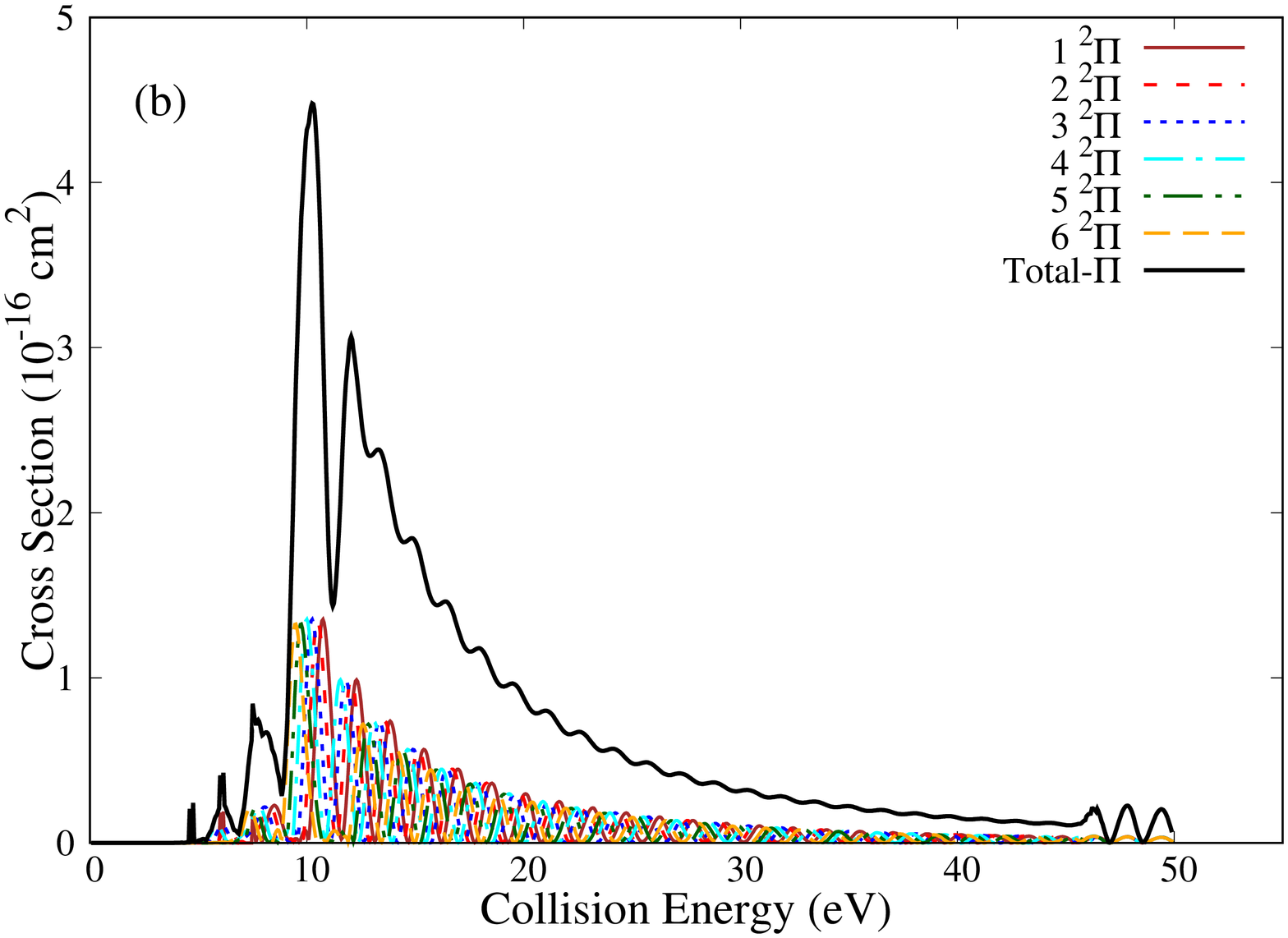}
\includegraphics[scale=0.35,angle=-90]{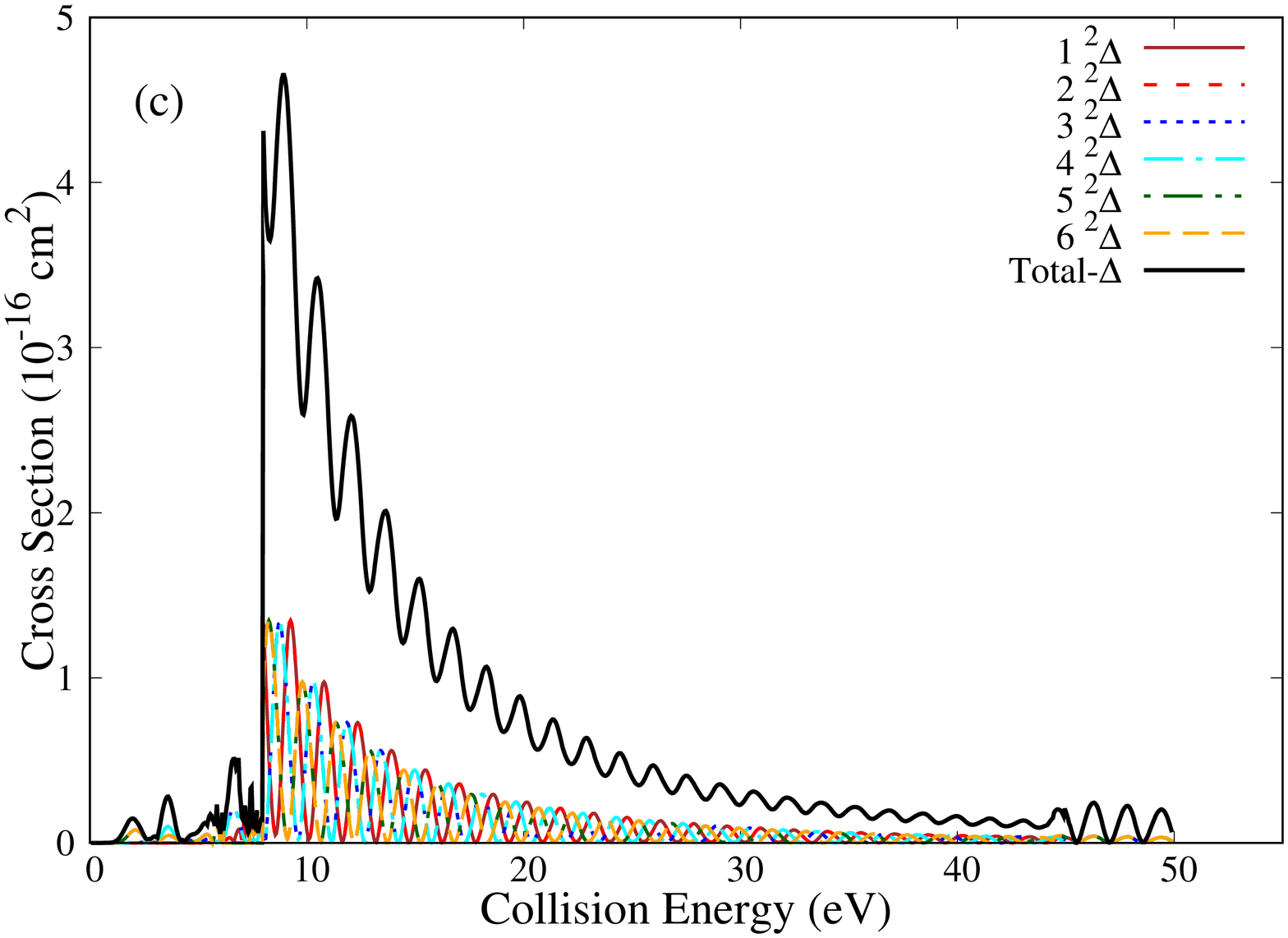}
\caption{\label{cross1} Calculated DR reaction cross section for $\mathrm{^4HeH^+}$ for the different symmetries, in the adiabatic representation.}
\end{figure}

The DR reaction cross section is computed by first propagating the wave packets on the 23 adiabatic potential  energy curves of $\mathrm{^4HeH}$. Rotational coupling between states of different symmetries are here taken into consideration. The  calculated partial cross sections for each of the states in the different symmetries are shown in figures~\ref{cross1}(a),~\ref{cross1}(b) and \ref{cross1}(c). The states for $^2\Pi$ and $^2\Delta$ symmetries are doubly degenerate~\cite{Larson15}. Thus, as seen in figure~\ref{cross1}(b) and figure~\ref{cross1}(c), two values of the cross sections are lying on top of each other. The threshold energy, for all symmetries in the adiabatic representation, is above six eV. Below this energy, the $\mathrm{HeH}$ resonant states are not dissociating into the neutral fragments. 
Propagating wave packets on the $\mathrm{HeH}$ resonant states  produces a DR reaction cross section that has some sharp peaks and oscillations. Some of these peaks can be explained by the shape resonances formed when the potential energy curve of a state has a barrier, in particular at low collision energies. The regular oscillations that are observed at above 10 eV are due to energy dependence of the electron capture probability~\cite{Motapon06,Roos08}.

\begin{figure}[!h]
\includegraphics[scale=0.35,angle=-90]{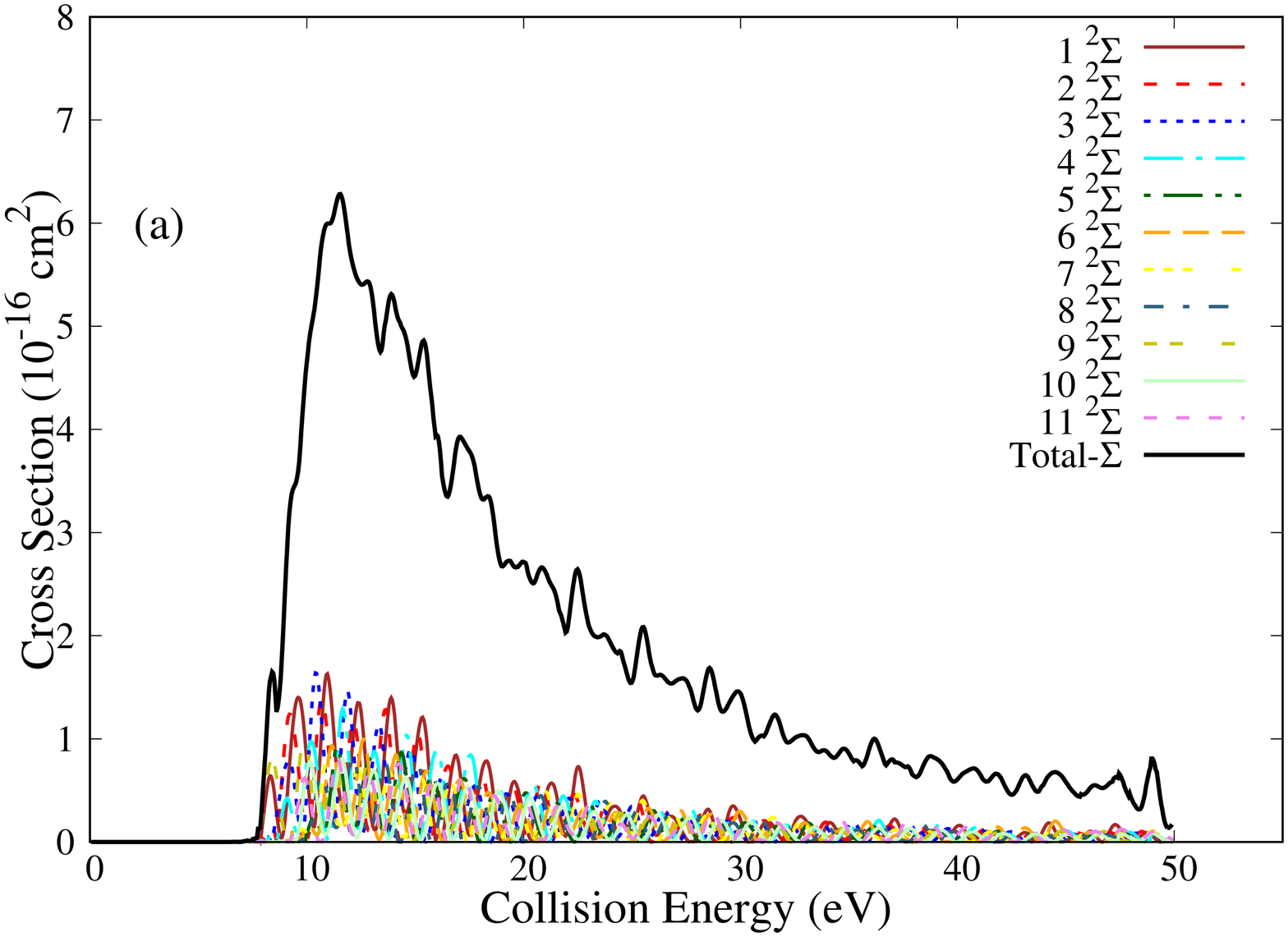}
\includegraphics[scale=0.35,angle=-90]{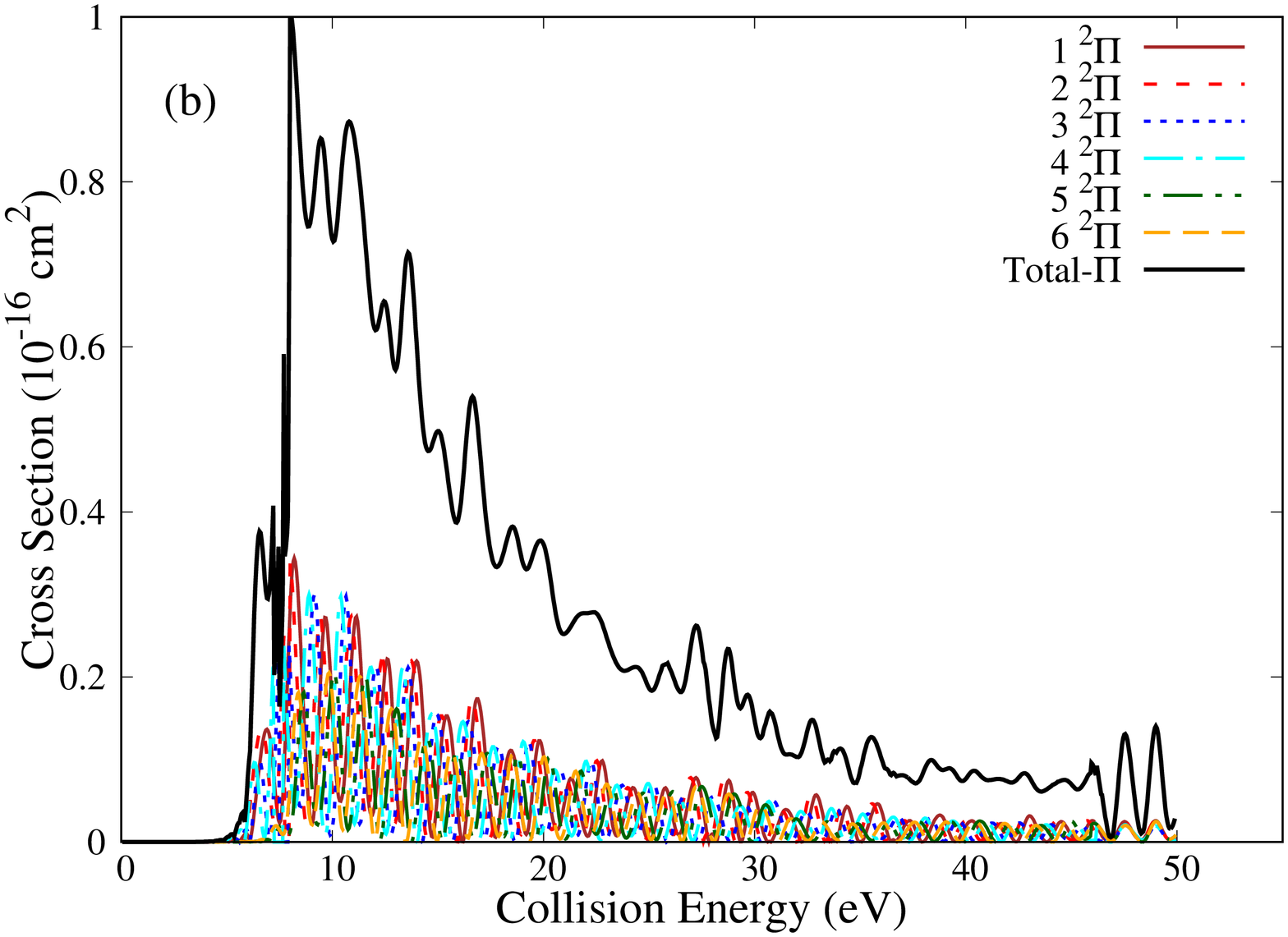}
\includegraphics[scale=0.35,angle=-90]{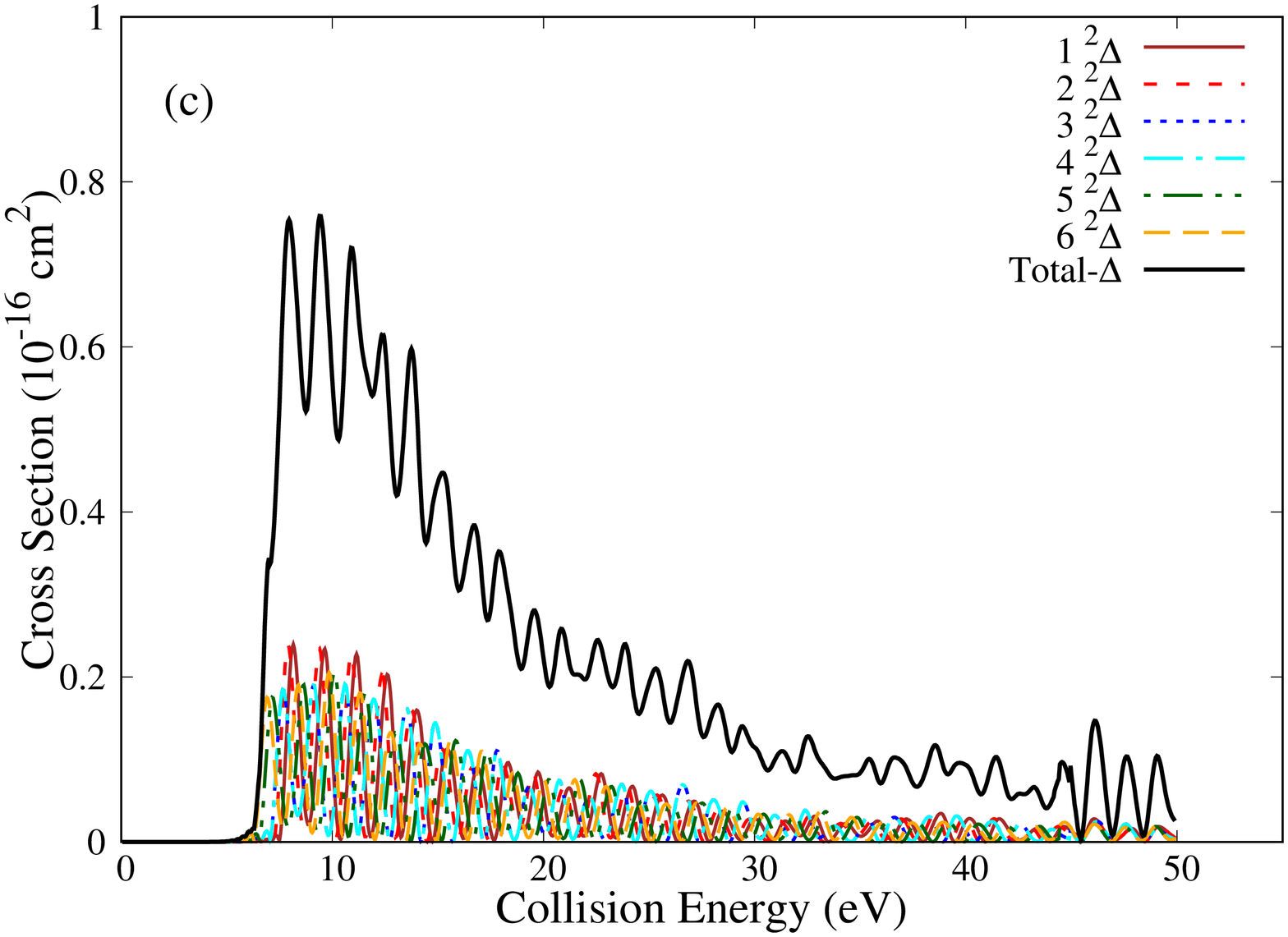}
\caption{\label{cross2} Calculated DR reaction cross section for $\mathrm{^4HeH^+}$ for the different symmetries, in the diabatic representation.}
\end{figure}

The DR reaction cross section is also computed in the diabatic representation. In this representation, the states have no non-adabatic couplings but are coupled by electronic couplings. The DR reaction cross section results are shown, for the three symmetries, in figures~\ref{cross2}(a),~\ref{cross2}(b) and \ref{cross2}(c). It should be noted that in this representation, the electronic states are not pure $\Sigma$, $\Pi$ or $\Delta$ symmetry~\cite{Larson16}. In this representation, the threshhold energy is above eight eV. 

\begin{figure}[!h]
\includegraphics[scale=0.35,angle=-90]{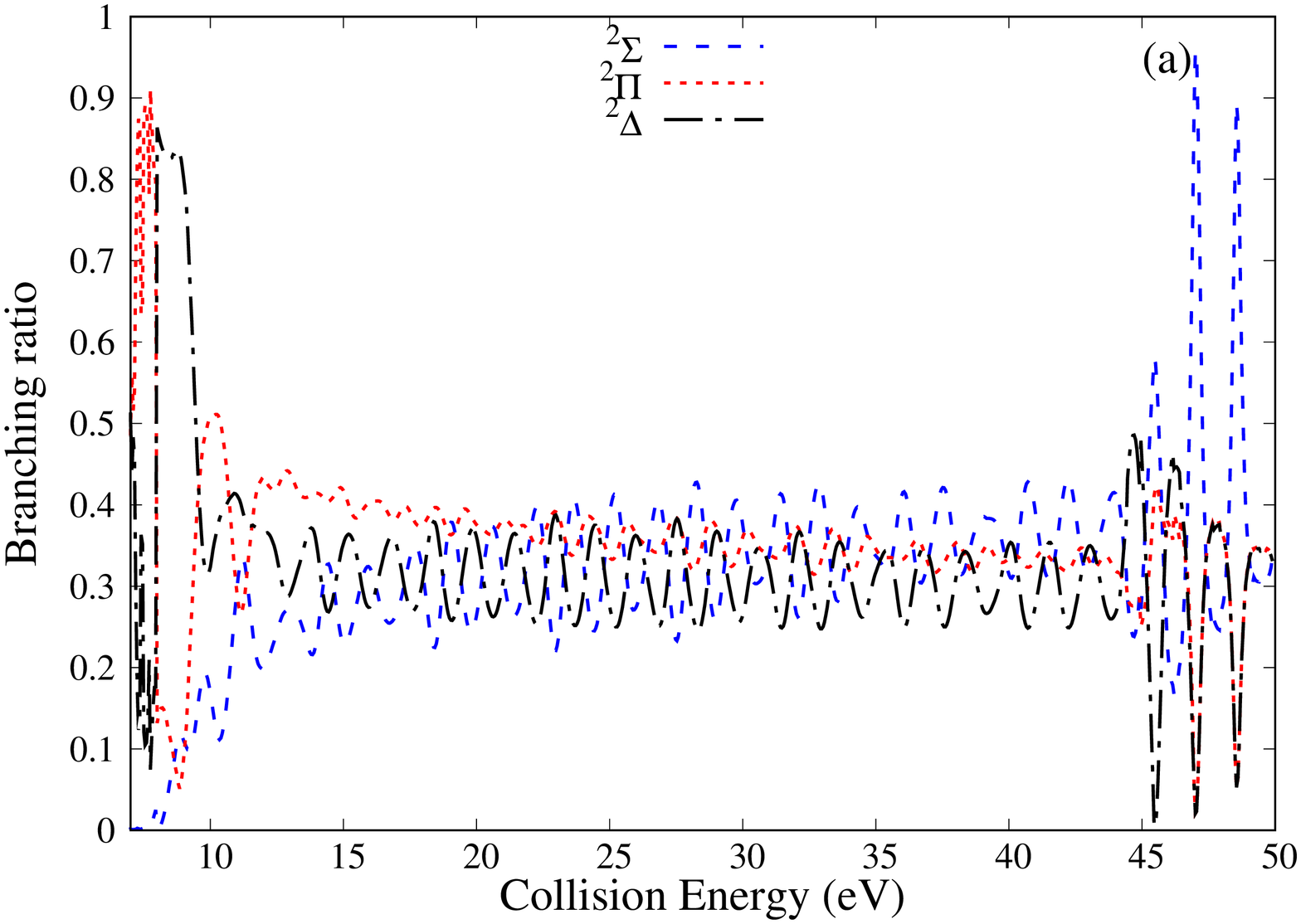}
\includegraphics[scale=0.35,angle=-90]{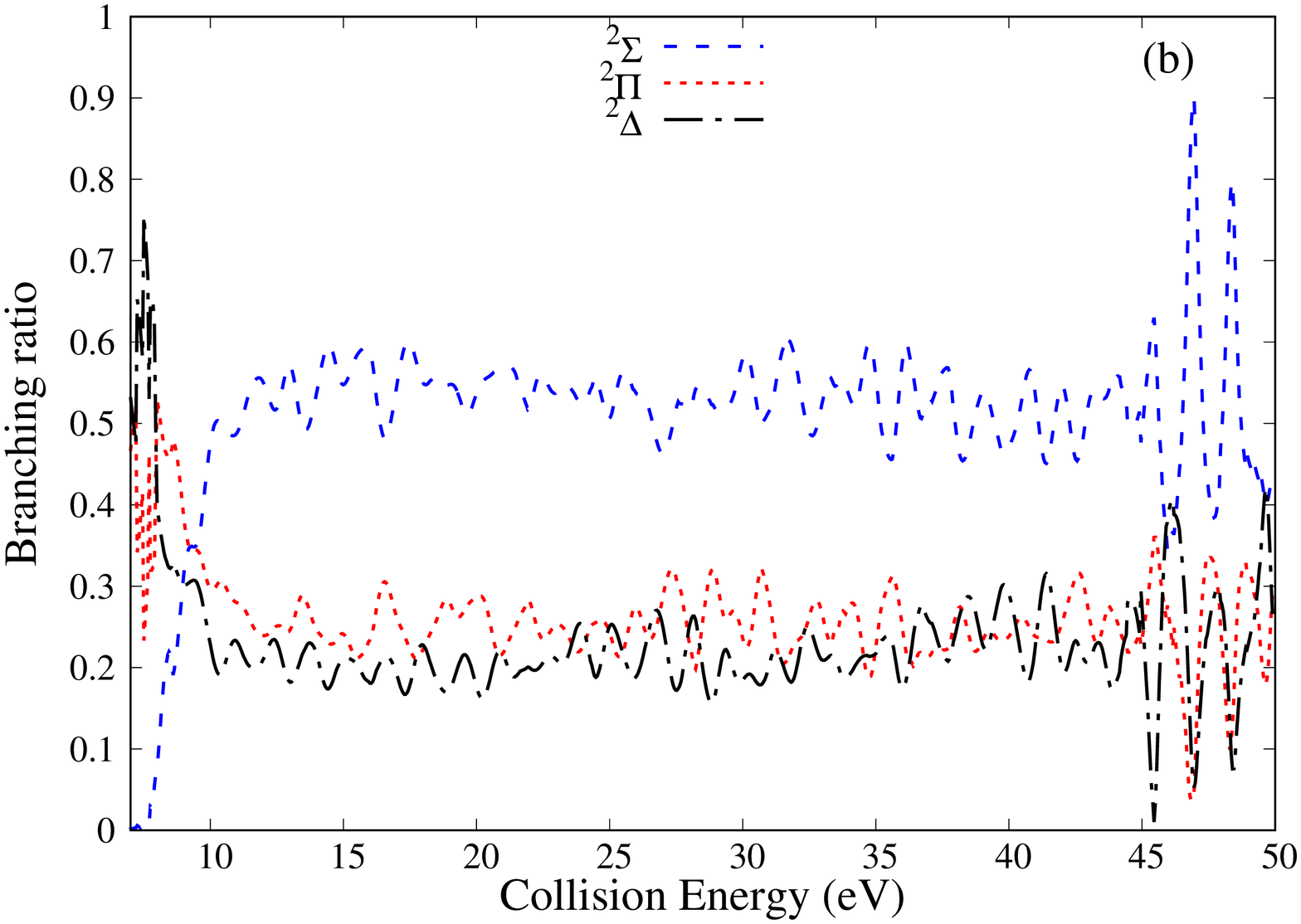}
\caption{\label{branch1} Symmetry contributions; showing a ratio of a contribution from each symmetry to the total DR reaction cross section in (a) adiabatic representation and (b) diabatic representation.}
\end{figure}

The DR reaction cross section contribution of electronic states for each symmetry to the reaction's total cross section is investigated in both representations. Branching ratios are used, which are ratios of the DR reaction total cross section from each symmetry to the total cross section. Figure~\ref{branch1} show the branching ratios for the adiabatic (a) and diabatic representations (b). In both representations, states of the $^2\Sigma$ symmetry are contributing more to the total cross section, for collision energies above 10 eV. Below 10 eV, states of the $^2\Pi$ symmetry are also contributing significantly. The $^2\Sigma$ states are contributing more to the cross section in the diabatic representation than in the adiabatic representation. States of the $^2\Delta$ symmetry contributes the lowest to the total cross section in the diabatic representations.

\subsection{ Study Isotope Effect in DR reaction of $\mathrm{HeH^+}$}

\begin{figure}[!h]
\includegraphics[scale=0.35,angle=-90]{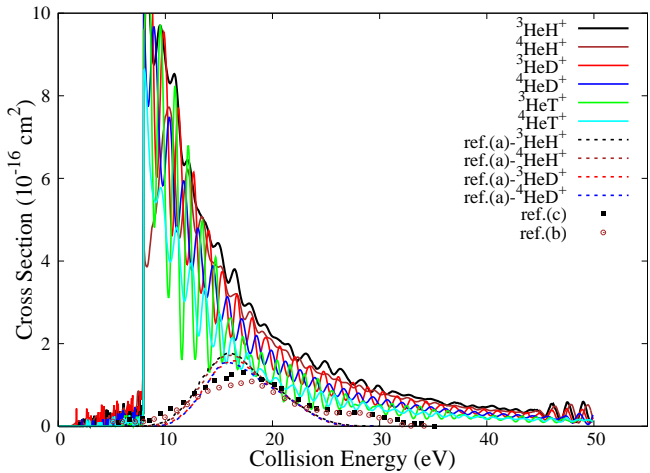}
\caption{\label{isot1} DR reaction total cross section for the $\mathrm{HeH^+}$ isotopologues. The results are compared with other theoretical results [ref.(a)~\cite{Larson98}], shown as lines. The comparisons with experimetal results [ref.(b)~\cite{Mowat95} and ref.(c)~\cite{Yousif89}], is shown as points.}
\end{figure}

The DR reaction cross section is also calculated for the $\mathrm{^3HeH^+}$, $\mathrm{^3HeD^+}$, $\mathrm{^3HeT^+}$, $\mathrm{^4HeD^+}$ and  $\mathrm{^4HeT^+}$ isotopologues. The study is carried out in the adiabatic representation. The cross sections are computed for all states in all symmetries and are shown in figures~\ref{isot1}. The total cross section results are compared with  the DR reaction cross sections that have been reported before from the theoretical~\cite{Larson98} and experimental results~\cite{Yousif89,Mowat95}. These results are only reported for some isotopologues, hence there are no comparisons for the DR reaction of $\mathrm{^3HeT^+}$ and $\mathrm{^4HeT^+}$. For all isotopologues, the electronic states in the $^2\Sigma^+$ symmetry are contributing more to the total cross section. An observable trend in the DR reaction cross section from the different isotopologues is the increase in the value to the total crossm section with the decrease in the reduced mass.

\subsection{RIP Formation Cross Section}
\begin{figure}[!h]
\includegraphics[scale=0.35,angle=-90]{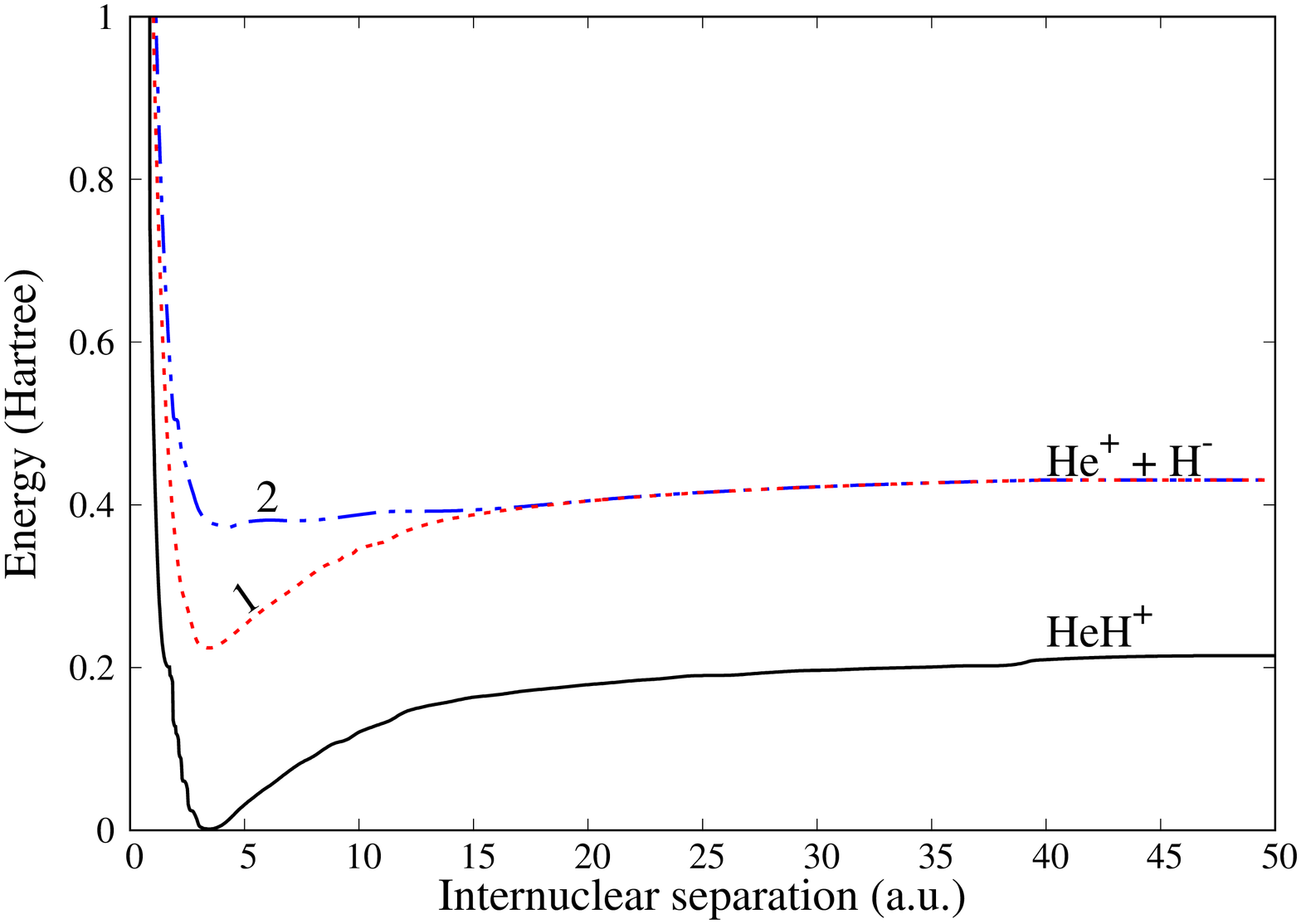}
\caption{\label{ip1} Potentail energy curves for the $\mathrm{He^+ + H^-}$ state and the ground state of $\mathrm{HeH^+}$.}
\end{figure}

For the ion-pair formation process, the calculation is carried out using two different methods: In \textbf{method 1}, the wave packets are propagated on an ion-pair potential energy curve obtained from the two-by-two adiabatic-to-diabatic transformation~\cite{Nkambule22}. In this approach, only two adiabatic states are assumed to be interacting near an avoided crossing and thus all other states are neglected in the transformation. Thus the states are transformed such that the ion-pair/covalent character is preserved~\cite{Hedberg14,Nkambule22}. The ion-pair state potential energy curve obtained using this approach is labelled 1 in figure~\ref{ip1}. In \textbf{method 2}, the wave-packets are propagated on a potential energy curve obtained from the strict adiabatic-to-diabatic transformation of 23 coupled states~\cite{Larson16}. Using this approach, the potential energy curves for the diabatic states will cross each other~\cite{Larson16}. However, the ion-pair and covalent character is preserved on each of the potential energy curves. The ion-pair character will be for the potential energy curve that is highest in energy at large internuclear distances. The potential energy curve for the electronic states obtained using this procedure is labelled 2 in figure~\ref{ip1}. These potential energy curves are all shifted such that the minimum energy of the ground state of the ion ($\mathrm{HeH^+}$) has its energy equal to zero.

\begin{figure}[!h]
\includegraphics[scale=0.35,angle=-90]{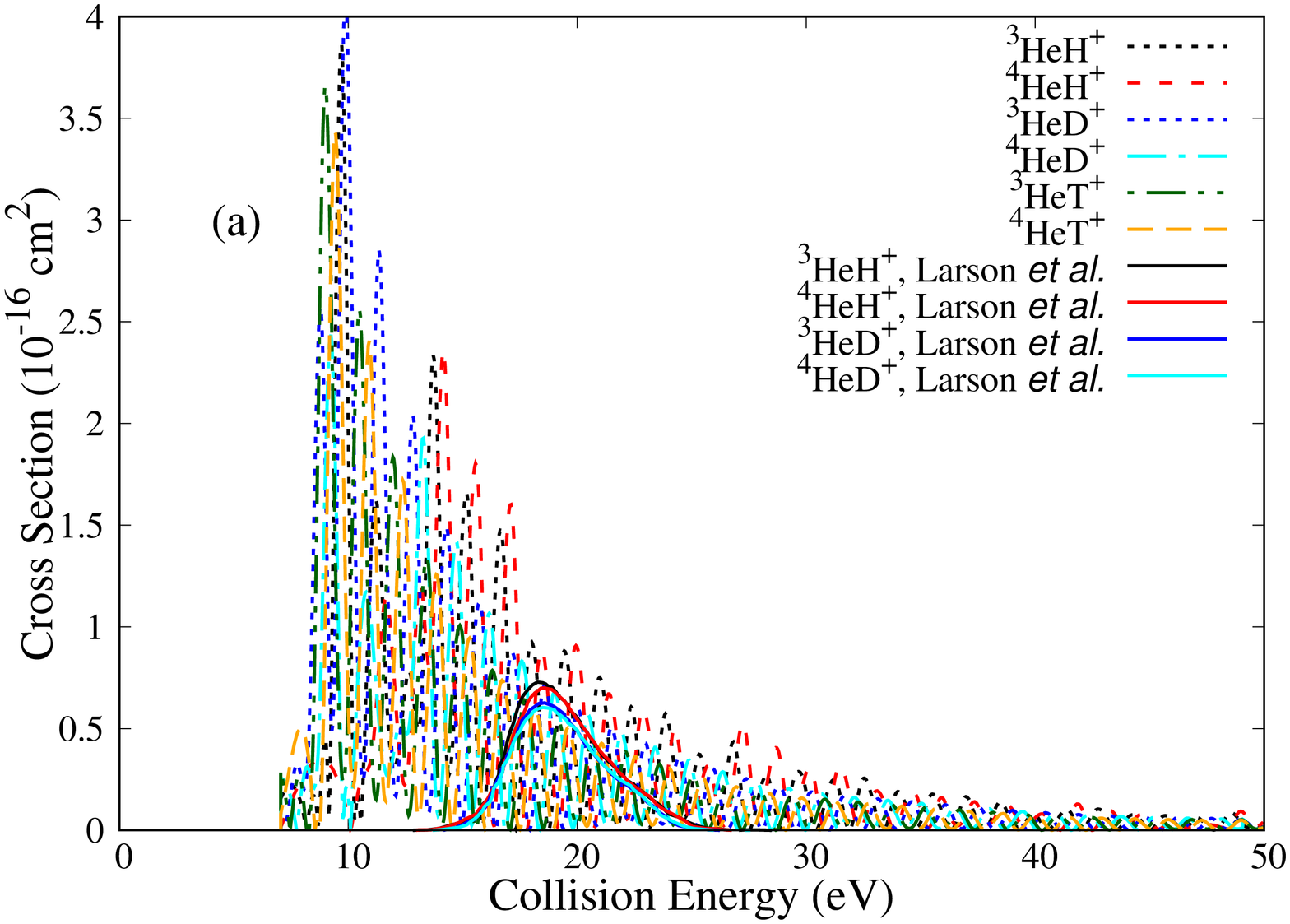}
\includegraphics[scale=0.35,angle=-90]{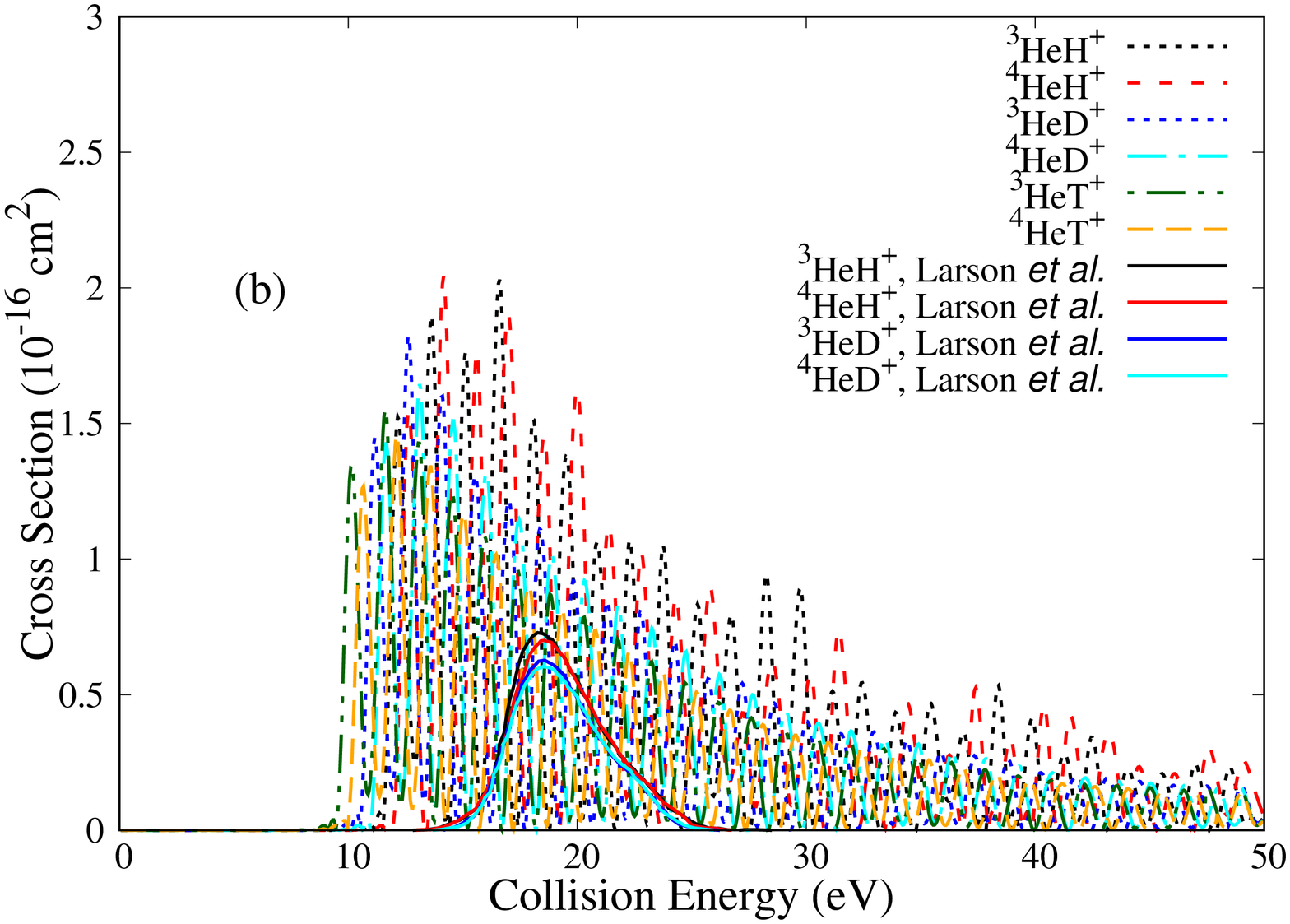}
\caption{\label{rip1} RIP formation total cross section for for different isotopes of $\mathrm{He}$ and $\mathrm{H}$ when using (a) \textbf{method 1} and (b) \textbf{method 2}.}
\end{figure}

The resonant ion-pair formation reaction total cross section for all different isotopologues obtained by using \textbf{method 1} and \textbf{method 2} are shown in figure~\ref{rip1}(a) and figure~\ref{rip1}(b), respectively. The results are compared with previous theoretical results~\cite{Larson98}. The current cross section is about twice as large as the previous results. This could be attributed to the treatment of all states as interacting, via non-adibatic couplings, and the inclusion of rotational couplings between states of different symmetries. The significant number of states included in this study compared to the previous study is also another factor. At low collision energies (below 12 eV) , the theoretical model used by Larson \textit{et al}~\cite{Larson98} yeilded a value close to zero. The current method is yielding a very large cross section in both procedures. In \textbf{method 1}, in the energy range $16$ eV to $22$ eV, the reaction cross section is almost equal with the previous results. However in \textbf{method 2}, the cross section is still larger, even in this range. The cross section, thereafter, decreases with collision energy. Other competing processes~\cite{Larson2014} could be significantly impacting the ion-pair formation process.

Cross section from collisions of lighter isotopes is bigger than cross sections from heavier isotopes. The same trend was observed previously~\cite{Larson98}. The ion-pair reaction cross section goes to zero at higher collision energies. 

\section{\label{concl}Conclusion}

The reaction cross section for the DR and RIP formation has been theoretically studied using time-dependant methods. The resulsts are showing a larger DR reaction cross section when comparing with previous results~\cite{Larson98,Mowat95,Yousif89}. For the theoretical results, the discrepancies are attributed to the difference in number of states that are included in the model. In current study 23 resonant states are included while in the previous study ony six states were included. The DR reaction cross section decreases, as the collision energy increases, in particular above 20 eV. This could be due to other competing processes~\cite{Larson2014} that start to be significant. The sharp peaks and oscillation in the cross section could be attributed to shape resonances formed by barriers in the potential energy curve landscape. At higher energies, the oscilations could be due to the fact the the electron capture probability is energy dependant~\cite{Motapon06,Roos08}. For the DR process, the cross section exhibits the prevalence of quantum tunneling in the adiabatic representation than in the diabatic representation.

The rotational couplings seem to be increasing the DR reaction flux to the states of $^2\Sigma$ symmetry as observed by the high branching section ratio in the diabatic representation. Although the $^2\Sigma$ states contribution is still larger at higher energies, their contribution is slightly lower than the $^2\Pi$ and $^2\Delta$ states in the adiabatic representation. The DR reaction cross section for species with a bigger reduced mass is lower than that from a lighter reduced mass.
In the RIP formation reaction cross section using \textbf{method 1}, the sharp peaks at low collision energies ($<10$ eV) could be caused by the shape of potential energy curve. RIP formation when using \textbf{method 1} gives a larger reaction cross at lower energies than \textbf{method 2}. For collision energies above 15 eV, \textbf{method 2} yields a larger reaction cross section.  
%% References

%%

\section*{Acknowledgemets}

We acknowledge Prof. \AA sa Larson from Stocholm University and Prof. Ann E. Orel from University of Carlifonia, Davis for their contribution with  the electronic structure and electron scattering data that was used in this work.
\section*{Declarations}

\subsection*{Data Availability Statement}
Data used in this work can be made available at request. Data Generously provided by Prof. \AA. Larson and Prof. A. E. Orel was analysed using our own FORTRAN codes. Derived data supporting the findings of this study are available from the corresponding author upon reasonable request.
\subsection*{Author Contribution}
All authors contributed to the study conception and design. Material preparation, data collection and analysis were performed by Sifiso M. Nkambule, Malibongwe Tsabedze, Oscar N. Mabuza and Mbuso K. Matfunjwa. The first draft of the manuscript was written by Sifiso M. Nkambule and all authors commented on previous versions of the manuscript. All authors read and approved the final manuscript.
\subsection*{Funding}
No funding was received for conducting this study.
\subsection*{Financial Interests}
The authors have no conflicts of interest to declare that are relevant to the content of this article.
%\begin{widetext}
%\nocite{*}
%\end{widetext}
\normalem
\pagebreak
\bibliography{references.bib}% Produces the bibliography via BibTeX.
\end{document}